\documentclass[aps, pra, twocolumn, 10pt, balancelastpage, amsmath, amssymb,superscriptaddress]{revtex4-2}

\usepackage[utf8x]{inputenc}
\usepackage[T1]{fontenc}
\usepackage{graphicx}
\usepackage{dsfont}
\usepackage{braket}
\usepackage{siunitx}
\usepackage{orcidlink}

\usepackage{hyperref}
\hypersetup{colorlinks = true, allcolors=blue, pageanchor}

\begin{document}

\title{Noise Effects on Diabatic Quantum Annealing Protocols}

\author{Giulia Salatino\,\orcidlink{0009-0005-6913-6920}}
\email{giulia.salatino@unina.it}
\affiliation{Scuola Superiore Meridionale, Largo San Marcellino 10, I-80138 Napoli, Italy}

\author{Maximilian Matzler\,\orcidlink{0009-0009-5833-9916}}
\affiliation{Institut f\"ur Theoretische Physik, Universit\"at Innsbruck, 6020 Innsbruck, Austria}
\affiliation{Institute for Quantum Optics and Quantum Information of the Austrian Academy of Sciences, 6020 Innsbruck, Austria
}

\author{Annarita Scocco\,\orcidlink{0000-0002-6920-5843}}
\affiliation{Scuola Superiore Meridionale, Largo San Marcellino 10, I-80138 Napoli, Italy}

\author{Procolo Lucignano\,\orcidlink{0000-0003-2784-8485}}
\author{Gianluca Passarelli\,\orcidlink{0000-0002-3292-0034}}
\email{gianluca.passarelli@unina.it}
\affiliation{Dipartimento di Fisica, Universit\`a di Napoli ``Federico II'', I-80126 Napoli, Italy}
    
\date{\today}

\begin{abstract}
    Diabatic quantum annealing aims to mitigate the challenges posed by small energy gaps and decoherence in quantum optimization by exploiting nonadiabatic transitions. In this paper, we compare the performance of two diabatic protocols in a realistic, dissipative setting using the maximum weighted independent set problem as a test case, showing that the potential advantages of diabatic protocols are strongly reduced in the presence of dissipation. Our results highlight the importance of evaluating quantum annealing strategies under practical conditions.
\end{abstract}

\maketitle
    
\section{Introduction}
\label{sec:intro}

Adiabatic quantum computing (AQC) and quantum annealing (QA) have emerged as promising paradigms, polynomially equivalent to gate-based quantum computation~\cite{aharonov2008aqcequivalence}, for solving complex optimization problems using quantum mechanics~\cite{finnila2014qa,kadowaki1998qa,farhi2000aqc,albash2018review}. Combinatorial optimization problems, which are prevalent in fields such as logistics, finance, material science, and more, are often difficult for classical algorithms due to their complexity and the exponential growth of the solution space with problem size~\cite{karp1972npcomplete,papadimitriou1998optimization}. AQC and QA offer a quantum approach to solve these problems by exploiting quantum tunneling to navigate the solution space more efficiently than classical methods~\cite{perdomo2012proteins,bian2013ramseynumbers,ronnow2014detectingquantumspeedup,rieffel2015operationalplanning,azinovic2017satfilters,mott2017higgs,li2017compbio,mandra2018speedup,jiang2018factorization,venturelli2019portfolio,smelyanskiy2020delocalized,zlokapa2021particletracking}. In these frameworks, a quantum system is initialized in the ground state of an easily solvable Hamiltonian and is then slowly evolved to a final Hamiltonian whose ground state encodes the solution to the problem of interest. The quantum adiabatic theorem is the basis for these approaches, ensuring that if the evolution is sufficiently slow, the system will remain in its ground state throughout the process, leading to an accurate solution~\cite{jansen2007adiabaticapproximation,morita2008foundation}.

However, the practical implementation of AQC and QA faces significant challenges~\cite{johnson2011manufactured,hauke2020implementations}. One of the primary issues is the presence of small energy gaps between the ground state and excited states during the evolution. According to the adiabatic theorem, the evolution time must be inversely proportional to the square of the minimum gap encountered along the path. As a result, when the gap is small, the required evolution time can become impractically long, making the system susceptible to decoherence and noise~\cite{albash2015decoherence} despite their intrinsic robustness~\cite{childs2001aqcrobustness}. These limitations compromise the efficiency and scalability of AQC and QA, since real-world quantum devices operate in noisy environments where prolonged operating times worsen the effects of decoherence~\cite{preskill2018nisq}. To address this, strategies that attempt to exploit decoherence have been put forward~\cite{amin2008thermallyassistedaqc,dickson2013thermallyassistedqa,arceci2017dissipativelz,smelyanskiy2017quantumdiffusion,passarelli2018dissipative}, such as pausing the annealing process to allow the system to relax into lower energy states~\cite{marshall2019pausing,passarelli2019pausing,albash2020comparing,chen2020whypausing,izquierdo2020ferromagnetically}, and reverse annealing~\cite{perdomo2011sombrero,lanting2014entanglementqa,albash2015entanglementqa,ohkuwa2018reverse,venturelli2019portfolio,yamashiro2019ara,ikeda2019nurse,passarelli2020reverse,grant2021portfolio,kumar2020fairsampling,chancellor2021searchrange,bando2022breakdown,passarelli2022ara}, where the annealing direction is temporarily reversed to escape local minima. However, in general, the most effective way to counteract decoherence is to reduce the annealing time as much as possible, beyond the adiabatic theorem, in the coherent limit~\cite{king2022coherent}.

Shortcuts to adiabaticity (STA)~\cite{gueryodelin2019sta} have emerged as a promising class of techniques to overcome the limitations imposed by the adiabatic theorem, enabling faster and more efficient quantum state preparation. Among these, counterdiabatic driving~\cite{demirplak2003adiabatic,delcampo2013stacd} has gained significant attention for suppressing transitions to excited states by introducing an auxiliary Hamiltonian. This approach is applicable to both continuous and digitized quantum annealing~\cite{farhi2014qaoa,wurtz2022counterdiabaticity,chandarana2022digitizedcounterdiabatic,vizzuso2024qaoacd,vizzuso2024gaps} and enables faster evolutions by mitigating the effects of small gaps and decoherence~\cite{delcampo2012adiabaticpassage,campbell2015stalmg,passarelli2020cd,passarelli2022cdopen,passarelli2023cra}. However, its practical implementation often requires detailed spectral knowledge of the problem Hamiltonian~\cite{berry2009transitionless}, and even approximate constructions are constrained by quantum speed limits~\cite{deffner2017qsl}. These challenges have spurred the development of more general STA methods, which, while less optimal, can be applied across a broader range of problems.  

Diabatic quantum annealing (DQA)~\cite{crosson2021diabatic} represents an alternative STA strategy that has attracted significant interest. Unlike traditional adiabatic approaches, DQA leverages controlled transitions to excited states, with the expectation that the system will return to the ground state by the end of the annealing process. This allows the system to bypass small gap regions that would otherwise require prohibitively long annealing times.  

One prominent proposal within this framework is nonstoquastic-DQA (NS-DQA), which employs nonstoquastic catalyst Hamiltonians to help diabatic transitions~\cite{hormozi2016nonstoquastic}. Under certain conditions, NS-DQA can achieve a diabatic path to the ground state even when the minimum gap closes exponentially~\cite{feinstein2024robustnessdiabatic}. However, the method requires precise tuning of parameters and remains sensitive to the specific annealing spectrum, highlighting the practical challenges of its implementation.  

Another recent DQA-inspired strategy is quantum quench dynamics, or the ``sweep-quench-sweep'' (SQS) protocol~\cite{lukin2024qqa}. In this approach, a quench is introduced between two quasi-adiabatic sweeps, enabling the system to bypass small gap regions more effectively. Experiments on programmable quantum simulators based on Rydberg atom arrays demonstrated significant improvements in ground-state fidelity, with orders-of-magnitude better performance compared to standard adiabatic protocols. The quench facilitates a macroscopic reconfiguration of the system, making SQS a compelling addition to the toolbox of STA techniques, particularly for larger and more complex quantum systems.

In this paper, our aim is to explore and compare the effectiveness of these two DQA protocols, NS-DQA and SQS, for solving a combinatorial optimization problem. Specifically, we focus on small instances of the maximum weighted independent set (MWIS) problem, a prototypical NP-hard optimization problem that is particularly well-suited for testing these quantum algorithms. Although NS-DQA and SQS may appear distinct, they both leverage the same core mechanism: inducing excitations in the quantum system before it encounters a small energy gap. This pre-gap excitation then results in population transfer back to the ground state once the small gap is encountered, thereby avoiding the need for a slow, fully adiabatic process.

However, the precision and control offered by these protocols differ significantly. NS-DQA is a tailor-made approach, carefully designed to ensure nearly complete population transfer back to the ground state at the transition, minimizing the likelihood of remaining in excited states. In contrast, SQS, which involves a quantum quench, is less precise and can lead to population leakage into other excited states due to the quench itself, though it still facilitates ground state transfer to some extent.

Decoherence, however, could drastically alter this scenario. By affecting the populations of the energy levels, decoherence could render the population transfer back to the ground state less effective as a shortcut to adiabaticity. This effect perturbs any diabatic protocol, since any small deviation from perfect population transfer due to decoherence could lead to significantly worse performance, as the finely tuned diabatic pathway could be disrupted. Understanding how these protocols perform in the presence of noise is therefore crucial to assessing their practical viability in real-world quantum devices.

The rest of this manuscript is organized as follows. In Sec.~\ref{sec:basics}, we review the basics of quantum annealing and provide a detailed description of both the NS-DQA and SQS protocols. In Sec.~\ref{sec:model}, we introduce the MWIS model and the specific instances we will study, following Refs.~\cite{feinstein2022xx,feinstein2024robustnessdiabatic}. In Sec.~\ref{sec:unitary}, we compare the performance of NS-DQA and SQS in terms of their ability to reach the ground state under unitary evolution. We assess how well these protocols can serve as shortcuts to adiabaticity, highlighting the strengths and limitations of each approach. In Sec.~\ref{sec:dissipative}, we examine the robustness of these protocols against decoherence by analyzing their open-system dynamics using a Markovian master equation. We investigate how the presence of environmental noise affects the effectiveness of NS-DQA and SQS. Finally, in Sec.~\ref{sec:conclusions}, we draw our conclusions.

\section{Adiabatic quantum computing and shortcuts to adiabaticity}
\label{sec:basics}

\subsection{Quantum annealing}
\label{subsec:qa}

Quantum annealing is a quantum optimization technique that aims to find the ground state of a problem Hamiltonian $ H_z $ by slowly evolving a quantum system from an initial Hamiltonian $ H_x $ that is easy to prepare. The system starts in the ground state of $ H_x $, typically a transverse field Hamiltonian, and is gradually transformed into the problem Hamiltonian $ H_z $ over a total annealing time $ T $. The Hamiltonian governing the system during this process is given by
\begin{equation}\label{eq:qa}
    \frac{H(t)}{J} = A(t) H_x + B(t) H_z,
\end{equation}
where $ A(t) $ and $ B(t) $ are time-dependent coefficients that control the evolution~\cite{kadowaki1998qa}. The value $ J $ fixes the energy scale, so that all the Hamiltonian terms in the following are assumed to be in units of it, while times will be expressed in units of $1/J$ with $\hbar = 1$. This will be assumed implicit in the following. In superconducting devices based on flux qubits, $J \sim \SI{1}{\giga\hertz}$~\cite{king2017dwavetechnicalsheet}, but in this paper we do not explicitly refer to any specific platform. The transverse field Hamiltonian $ H_x $ is often expressed as
\begin{equation}
    H_x = -\sum_{i=1}^N \sigma_i^x,
    \label{eq:transverse_field}
\end{equation}
where $ \sigma_i^x $ are the Pauli $ x $-matrices acting on the $ i $-th qubit, and $ N $ is the total number of qubits. The problem Hamiltonian $ H_z $ is typically chosen to be diagonal in the computational basis:
\begin{equation}
    H_z = \sum_{i=1}^N h_i \sigma_i^z + \sum_{i<j} J_{ij} \sigma_i^z \sigma_j^z,
\end{equation}
where $ h_i $ are local fields, $ J_{ij} $ are the coupling constants between qubits $ i $ and $ j $, and $ \sigma_i^z $ are the Pauli $ z $-matrices. Setting the local fields and the weights appropriately allows representing any combinatorial optimization problem in this form~\cite{lucas2014np}. 

In standard QA, the annealing schedules $ A(t) $ and $ B(t) $ are typically chosen to be linear functions of time:
\begin{equation}\label{eq:schedules}
    A(t) = 1 - \frac{t}{T}, \quad B(t) = \frac{t}{T}, \quad 0 \leq t \leq T.
\end{equation}
More optimized schedules could also be considered to improve performances~\cite{roland2002search,susa2021variationalschedule,matsuura2021variationalschedules,chen2022montecarlonnschedules,wauters2022graddescversusmcschedules,khezri2022customschedules,hegde2022genetic,hegde2023deeplearning}. At the beginning of the annealing process, $ A(t = 0) = 1 $ and $ B(t = 0) = 0 $, so the system is fully governed by the transverse field Hamiltonian $ H_x $, whose ground state is a uniform superposition of all computational basis states. As time progresses, $ A(t) $ decreases and $ B(t) $ increases, causing the system to evolve gradually from the ground state of $ H_x $ toward the ground state of $ H_z $. At the end of the annealing process, $ A(t = T) = 0 $ and $ B(t = T) = 1 $, leaving the system fully governed by $ H_z $. In the following, we will always assume $ A(t) = 1 - B(t) $ and use $ B(t) $ as the only annealing schedule.

If the evolution is sufficiently slow, the adiabatic theorem ensures that the system remains in its instantaneous ground state and ends in the ground state of $ H_z $, which encodes the solution to the optimization problem~\cite{jansen2007adiabaticapproximation}. The rate of evolution is constrained by the minimum energy gap, $ \Delta_{\text{min}} = \min_t \Delta(t) $, where $ \Delta(t) = E_1(t) - E_0(t) $ is the energy difference between the first excited state and the ground state of $ H(t) $. A smaller $ \Delta_{\text{min}} $ translates into a slower evolution to avoid transitions to excited states.

This minimum gap often occurs at a critical point where the system's ground state undergoes significant structural changes~\cite{sachdev2000qpt}. Fast evolution near this point can induce nonadiabatic transitions, resulting in a final state that does not solve the optimization problem. Thus, the total annealing time $ T $ must scale as $ T \propto \Delta_{\text{min}}^{-2} $ to ensure adiabaticity. For complex problems, where $ \Delta_{\text{min}} $ may become exponentially small~\cite{tanaka2017spinglass}, this scaling can render the required annealing time impractically long.

\subsection{Nonstoquastic diabatic quantum annealing}
\label{subsec:dqa}

Nonstoquastic diabatic quantum annealing extends the standard quantum annealing framework by introducing an additional Hamiltonian term, $ H_c $, which acts as a catalyst to facilitate diabatic transitions~\cite{feinstein2022xx}. The modified Hamiltonian during the annealing process is given by
\begin{equation}\label{eq:dqa}
    H_{\text{DQA}}(t) = A(t) H_x + B(t) H_z + C(t) H_c,
\end{equation}
where $ C(t) = A(t)B(t) $ is another time-dependent coefficient that controls the evolution of the system. The introduction of $ H_c $ is designed to generate an additional energy gap, $ \Delta_\text{c} $, which is comparable in size to the minimum gap, $ \Delta_{\text{min}} $, encountered in the standard annealing process. This additional gap plays a crucial role in enhancing the overall success probability of reaching the ground state.

As the system approaches the point where the minimum gap $ \Delta_{\text{min}} $ occurs, a diabatic transition to the first excited state becomes likely if the annealing time $ T $ is too short to maintain adiabaticity. The probability of the system undergoing this transition is given by the Landau-Zener (LZ) formula~\cite{landau1932crossings,zener1932crossings}
\begin{equation}\label{eq:lz1}
    P_{\text{excited}} \sim P_0 \exp\left(-T \Delta_{\text{min}}^2\right),
\end{equation}
where the ground-state population before the gap can be assumed to be $ P_0 \sim 1$. To counteract this, the additional term $ H_c $ is introduced to create a second gap $ \Delta_\text{c} $, positioned later in the annealing process. This second avoided crossing allows the system to relax back to the ground state. The effectiveness of this population transfer depends on the overlap between the first excited state and the ground state at $ \Delta_\text{c} $, with the probability of successfully returning to the ground state expressed as before via the LZ formula
\begin{equation}
    P_{\text{ground}} \sim P_{\text{excited}} \exp\left(-T \Delta_\text{c}^2\right),
\end{equation}
where we are assuming that the excited state is populated with a probability $ P_{\text{excited}} $ given by Eq.~\eqref{eq:lz1} (i.\,e., we assume no other transitions have occurred in-between). Ideally, the system reaches the end of the annealing process in the ground state of $ H_z $. The success of NS-DQA hinges on the careful design of the catalyst term $ H_c $, which must be specifically tailored to the problem's spectral properties. This ensures that $ \Delta_\text{c} $ is small enough to enable almost complete population transfer back to the ground state, minimizing the risk of population leakage into other excited states and maximizing the overall success of the annealing process. 

\begin{figure}[t]
    \centering
    \includegraphics[width = \columnwidth]{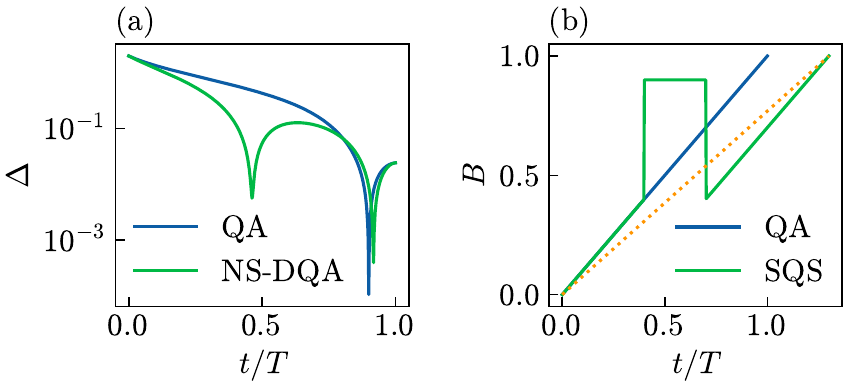}
    \caption{(a) Sketched comparison of the typical structure of the instantaneous gap $\Delta(t) = E_1(t) - E_0(t)$ in quantum annealing and NS-DQA. In the latter case, there is a second gap induced by the extra term in the Hamiltonian of Eq.~\eqref{eq:dqa}. (b) Comparison of the annealing function $ B(t) $ of quantum annealing and SQS. The dotted line represents a standard quantum annealing schedule lasting $ T' = T + \Delta T_{\text{q}} $.}
    \label{fig:qa-dqa-SQS}
\end{figure}

Notice that the role of $ \Delta_{\text{min}} $ and $ \Delta_\text{c} $ could be swapped without changing the scope of the algorithm, so $\Delta_\text{c}$ can also be induced by the catalyst \textit{before} $\Delta_{\text{min}}$: The system would get excited at $\Delta_\text{c}$, and later relax back to the ground state at $\Delta_\text{min}$. This scenario is depicted in Fig.~\ref{fig:qa-dqa-SQS}(a).

\subsection{Sweep-quench-sweep}
\label{sec:SQS}

In the sweep-quench-sweep protocol, the standard linear annealing schedule is modified to include a quench that disrupts the adiabatic evolution of the system at a conveniently chosen point of the dynamics. The Hamiltonian during this process remains as in Eq.~\eqref{eq:qa}, with $ A(t) $ and $ B(t) $ controlling the transition from the initial to the problem Hamiltonian equal to the unassisted case. 

Initially, the system evolves quasi-adiabatically, with $ A(t) $ decreasing and $ B(t) $ increasing gradually as in Eq.~\eqref{eq:schedules}. However, as the system approaches the critical region where the minimum gap $ \Delta_{\text{min}} $ is expected, a sudden quench is applied at $t = t_{\text{q}}$. During this quench, the annealing parameters $ A(t) $ and $ B(t) $ are rapidly adjusted,
\begin{equation}
A(t_{\text{q}}) \rightarrow A_{\text{q}}, \quad B(t_{\text{q}}) \rightarrow B_{\text{q}},
\end{equation}
causing the system to deviate from its adiabatic path and inducing nonadiabatic transitions. This quench mixes the ground state with excited states, particularly the first excited state, thereby creating a superposition that modifies the population distribution across these states~\cite{lukin2024qqa}.

The quench itself has a finite duration, during which the system evolves under the altered parameters $ A_{\text{q}} $ and $ B_{\text{q}} $. Since we assume $ A(t) = 1 - B(t) $, also $A_{\text{q}} = 1 - B_{\text{q}}$. A typical SQS protocol for $ B(t) $ is shown in Fig.~\ref{fig:qa-dqa-SQS}(b). The duration of the quench, $ \Delta T_{\text{q}} $, as well as the timing of its onset, $ t_{\text{q}} $, and the specific value of the quenched parameter, $ B_{\text{q}} $, are critical and must be carefully optimized to maximize the performance of the shortcut. 

After the quench, the parameters are set back to the values they had before the quench and the annealing process resumes from there, but now with a different state composition due to the previous mixing. This nonadiabatic interaction allows some of the excited population to relax back into the ground state, effectively using the dynamics induced by the quench to enhance the probability of ending in the ground state. 

More precisely, we use the following parametrization of the annealing schedule $B(t)$ in the SQS protocol:
\begin{equation}
    \begin{cases}
        B(t) = t/T & \text{if $0 \le t < t_\text{q}$;}\\[1ex]
        B(t) = B_\text{q} & \text{if $t_\text{q} \le t < t_\text{q} + \Delta T_\text{q}$;}\\[1ex]
        B(t) = \displaystyle\frac{t - \Delta T_\text{q}}{T} & \text{if $t_\text{q} + \Delta T_\text{q} \le t \le T + \Delta T_\text{q}$.} 
    \end{cases}
\end{equation}
In this protocol, the total annealing time is $ T' = T + \Delta T_{\text{q}} $. Therefore, to ensure a fair comparison, we will compare SQS against NS-DQA and standard quantum annealing with the same total duration $ T' $, see the dotted line in Fig.~\ref{fig:qa-dqa-SQS}(b).

\section{Model}
\label{sec:model}

The maximum weighted independent set is an NP-hard problem that has attracted much attention in the combinatorial optimization community~\cite{lamm2018exactlysolvingmaximumweight}. Given a graph $G=(V,E,\omega)$, with $\omega$ a weight function $\omega:V\rightarrow\mathbb{R}^+$, the goal of the problem is to find the maximally weighted set of vertices $G_0\in V$ such that no vertices are adjacent to each other. The MWIS problem can be naturally mapped to an Ising model, making it suitable for quantum annealing. In its Ising model encoding, the MWIS problem is represented such that each vertex in the graph corresponds to a spin. The possible sets of vertices are mapped to the computational basis states, where a spin in the up state indicates inclusion of the vertex in the set $G_0$, and a spin in the down state indicates exclusion. Within this encoding, flipping a spin corresponds to adding or removing a vertex from the independent set. The problem Hamiltonian has the form~\cite{feinstein2024robustnessdiabatic,feinstein2022xx}
\begin{equation}
    H_z=\sum_{i\in V}\left(c_i j_{zz}-2\omega_i\right)\sigma_i^z+\sum_{(i,j)\in E}j_{zz}\sigma_i^z\sigma_j^z
    \label{eq:ham_mwis}
\end{equation}
where $\sigma_i^z$ is the Pauli-$Z$ operator acting on vertex $\nu_i\in V$, $c_i$ is the connectivity and $\omega_i$ is the weight of the vertex. The constant $j_{zz}$ is a parameter that sets the strength of the antiferromagnetic coupling between connected vertices, enforcing the independent set condition. The term $c_i$ represents the connectivity of vertex $i$, defined as its degree, i.\,e., the number of edges connecting it to neighboring vertices. Details about the derivation of this Hamiltonian can be found in Appendix~\ref{sec:app_hamiltonian}. 

Fig.~\ref{fig:mwis}(a) shows a sketch of the Ising formulation of the MWIS problem on a simple graph of five vertices. Even if the ground state of the Hamiltonian is easy to compute analytically, as $\bigotimes_{i\in G_0}\ket{\uparrow_i}\bigotimes_{i\in G_1}\ket{\downarrow_i}$, this does not imply that it is easy to retrieve it through a quantum annealing process.

In this problem the size of the system is fixed to be odd, and the two subgraphs $G_0$ and $G_1$ have $n_0 = (N-1)/2$ and $n_1 =(N+1)/2$ spins, respectively. We then set the connectivities $c_1 = n_0$ for all the sites in $G_1$ and $c_0=n_1$ for all the sites in $G_0$.

\begin{figure}[t]
    \centering
    \includegraphics[width=\columnwidth]{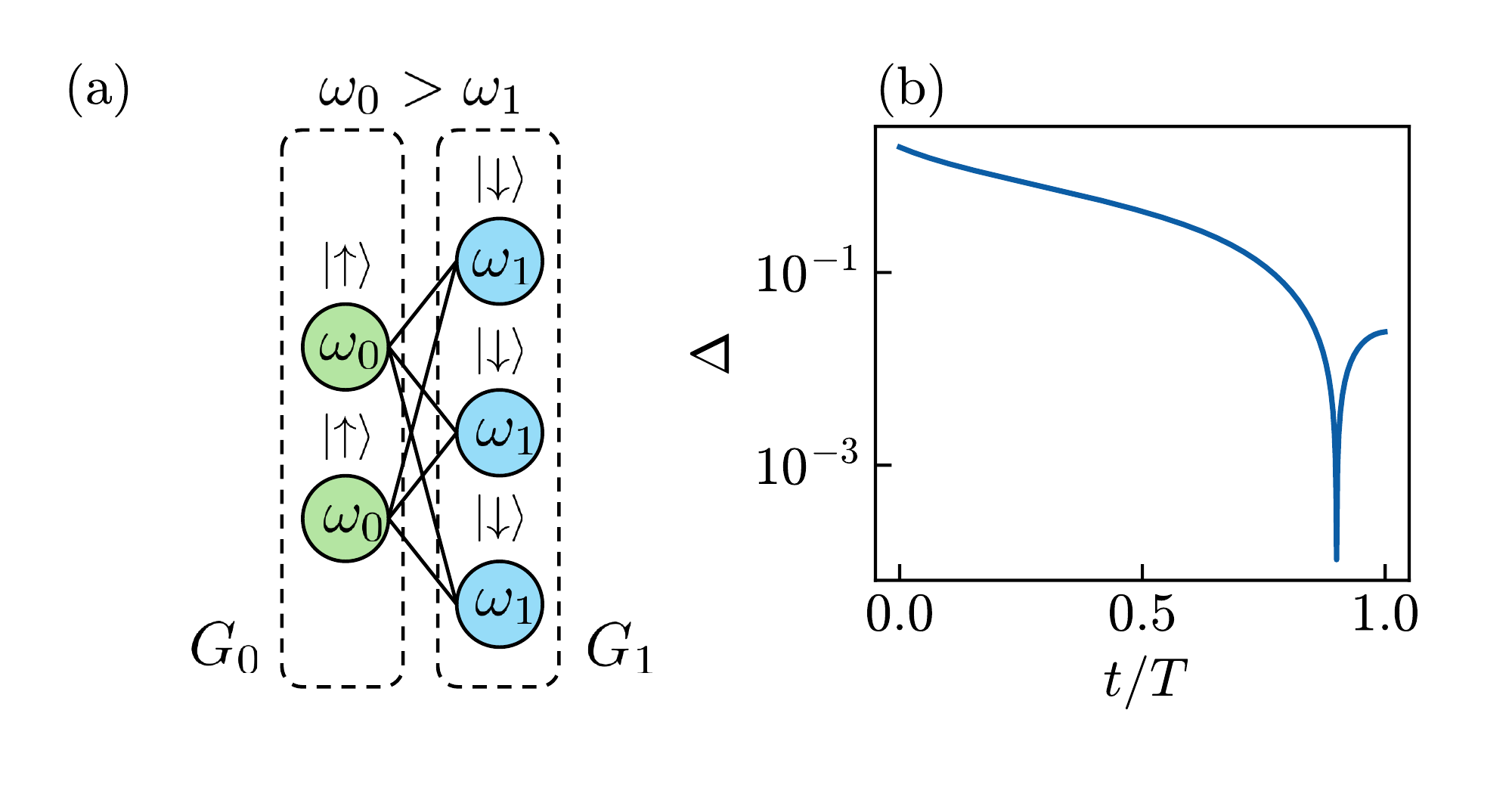}
    \caption{(a) Sketch of the Ising formulation of the MWIS problem on a simple graph. (b) Instantaneous gap of the annealing process. Energies are implicitly in units of $J$.}
    \label{fig:mwis}
\end{figure}

In the annealing process, the driving Hamiltonian is chosen as the homogeneous local transverse field of Eq.~\eqref{eq:transverse_field}. In the context of the MWIS problem, perturbative crossings can emerge during the annealing process, leading to avoided level crossings that play a crucial role in determining the system's evolution. To understand the formation of these avoided crossings, one can consider the driver Hamiltonian $H_x$ as a perturbation to the problem Hamiltonian $H_z$. In this picture, the energies of the states in the vicinity of a crossing are adjusted due to the perturbative influence of $H_x$. When the perturbation causes the energy of a low-lying excited state to approach that of the ground state, an avoided crossing is formed. The Hamiltonian near the crossing can be approximated as a two-level system, where the perturbative effect introduces a finite gap between the states. We refer to Refs.~\cite{feinstein2022xx, feinstein2024robustnessdiabatic} for further details on the formations of the avoided crossings.
Due to the latter, during the annealing process, the gap between the ground and first-excited state closes exponentially in the system size~\cite{feinstein2024robustnessdiabatic}, in a way that strictly depends on the setting of the parameters of the model.

Suitably setting the parameters of the annealing protocol, the gap $\Delta$ between the ground and the first excited state closes almost at the end of the annealing process. The discussion on the parameters setting can be found in Appendix~\ref{sec:app_hamiltonian}. Fig.~\ref{fig:mwis}(b) shows the gap closure with our choice of parameters, for $N = 5$. Due to the exponential closing of the gap, the success of the annealing protocol is drastically compromised already for very small sizes.

\section{Unitary dynamics}
\label{sec:unitary}

In this section we begin by recalling the characteristics of a standard quantum annealing process with the Hamiltonian of the MWIS problem. This proves the need to move to diabatic quantum annealing protocols, which we will analyze below. We quantify the success of the protocol through the analysis of the fidelity, defined as $F(t,T)=|\braket{\psi(t,T)|E_i(t/T)}|^2$ --- where $\ket{\psi(t,T)}$ is the evolved state and $\ket{E_i(t/T)}$ is the instantaneous eigenstate related to the $i$-th energy level $E_i$ of the annealing protocol. The fidelity depends on both $t$ and $T$, because the latter time determines the speed of the annealing process. Nevertheless, for brevity, we will hereafter refer to $F(t)$ for the instantaneous fidelity, namely the fidelity as a function of $t$ --- given a certain value of $T$ ---, $F(T)$ for the final fidelity as a function of the annealing time $T$, and, finally, $F(T;N)$ for the final fidelity at a certain value of $T$, as a function of the size $N$ of the system.

\subsection{Standard Quantum Annealing}
\begin{figure}
    \centering
    \includegraphics[width=\columnwidth]{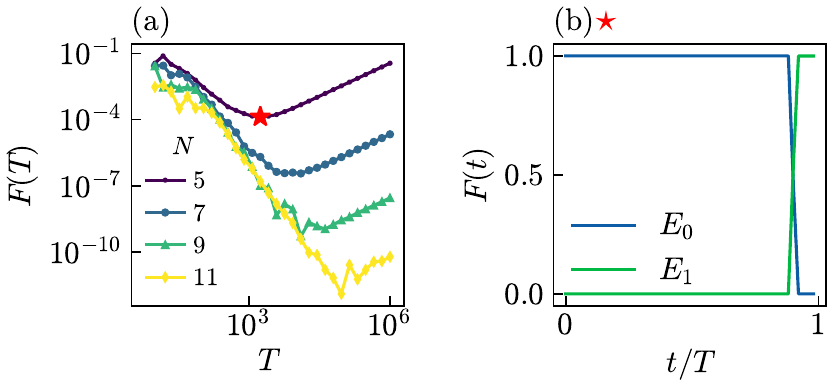}
    \caption{(a) Final-state fidelity $F(T)$ as a function of the annealing time $T$ for $N=5,7,9,11$. The red star represents the annealing time $T=2000$, whose dynamics is shown in panel (b). Times are implicitly in units of $J^{-1}$. (b) Instantaneous-state fidelity $F(t)$ as a function of $t$ for a choice of $T=2000$ for $N=5$.}
    \label{fig:fidelities_mwis}
\end{figure}

Let us first analyze some features of the  standard annealing process. Fig.~\ref{fig:fidelities_mwis}(a) shows the behavior of the final-state fidelity $F(T)$ as a function of the annealing time $T$ for $N=5,7,9,11$ ---  we do not investigate larger sizes since we already see fidelities dropping below machine precision. In it, we have considered a range of values of $T$ that spans from $T=10$ to $T\simeq10^6$. The plot demonstrates that for small values of annealing time ($\sim 10$), the final-state fidelity is different from zero. In particular, we can see that the final-state fidelity is equal to $F(T\sim10) \simeq 1/2^N$. This is due to the fact that the dynamics is so short in time that the state almost does not evolve, in fact remaining in the initial state of the protocol $\ket{\psi_0}=\bigotimes_{i=1}^N\ket{+}_i$. This leads to a probability of about $1/2^N$ of finding the state in the ground state of the problem Hamiltonian. On the other hand, for higher values of annealing time ($T = 10^2 \div 10^3$) the final fidelity tends to decrease, reaches a minimum value, and then starts growing again with the onset of the regime of validity of the adiabatic theorem.  Hence, increasing the annealing duration by increasing $T$ does not improve the success of the protocol until an unfeasibly large annealing time that grows exponentially with the size of the system~\cite{feinstein2022xx}. This is because the gap closure, caused by the avoided level crossing, leads to a complete depletion of the zero-energy level that only an exponentially-growing annealing  time, namely an extremely slow evolution, can prevent.

In Fig.~\ref{fig:fidelities_mwis}(b) we focus on a $5$-spin system and see the behavior of the fidelity with respect to the instantaneous ground state and the first excited state for an annealing process with $T=2000$, corresponding to the minimum fidelity observed in Fig.~\ref{fig:fidelities_mwis}(a) for that size, marked with a red star. As soon as the gap closes, the probability of finding the system in the ground state sharply drops, and the system completely occupies the first excited state, which naturally leads to the poor success of the adiabatic annealing protocol. In fact, in this case the minimum gap is of the order of $\mathcal{O}(10^{-4})$, so we are in the regime $T \ll \Delta^{-2}$. 

The results shown are derived by numerically solving the time-dependent Schr\"odinger equation using the Python library QuTiP~\cite{lambert2024qutip5}. This introductory result clarifies the importance of considering diabatic mechanisms, which we will analyze below.

\subsection{Diabatic Quantum Annealing}
In the following, we conduct a comparison between NS-DQA and SQS, when the system undergoes unitary dynamics. While nonstoquastic diabatic quantum annealing has been extensively tested in solving the optimization problem under consideration in Refs.~\cite{feinstein2024robustnessdiabatic,feinstein2022xx}, it is interesting to apply the SQS protocol to the same problem and assess its effectiveness. To this end, we start from the well-established model, setting its parameters in accordance with what has been widely studied in the literature and simply recall the salient features of the unitary diabatic annealing protocol. In contrast, for the SQS protocol, a preliminary optimization of the parameters will be required before testing its effectiveness and comparing it to that of the diabatic protocol. The optimization process is reported in Appendix~\ref{sec:sqs_optimization}. 

\begin{figure}[t]
    \centering
    \includegraphics[width=\columnwidth]{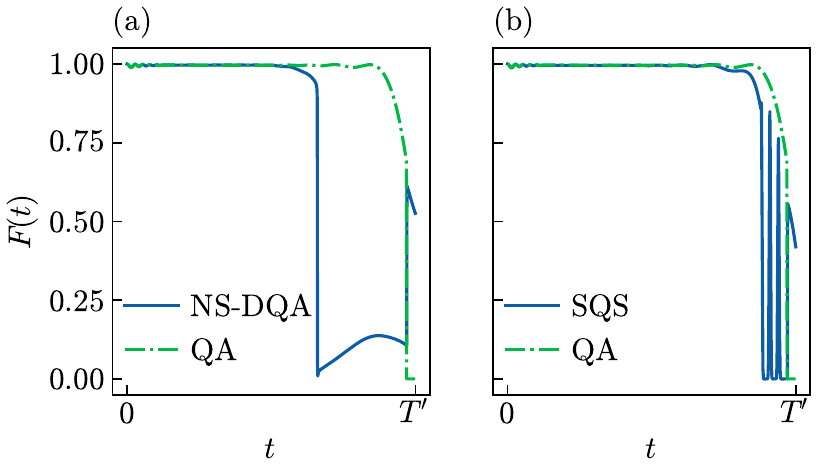}
    \caption{Instantaneous fidelity $F(t)$ for a system of $N = 11$ spins for an evolution of $T' = T + \Delta T_{\text{q}}$, where $ T = 100$ and $\Delta T_{\text{q}} = 8$. (a) Fidelity obtained using QA (green dot-dashed line) and the NS-DQA protocol (blue solid line).  (b) Fidelity obtained using QA (green dot-dashed line) and SQS (blue solid line). Times are implicitly in units of $J^{-1}$.}
    \label{fig:comp_dqa_sqs_unitary}
\end{figure}

In the following analysis, we will study a range of annealing times $T$ from $T = 100$ to $T \simeq 3000$, exploring different regimes of non-adiabaticity so as to extensively characterize the two different methods. We will furthermore repeat the same analysis for increasing sizes of the system, to perform a finite-size scaling. To avoid the exponential growth of the Hilbert space and explore bigger system sizes, we exploit the symmetries in the Hamiltonian. Indeed, the latter is invariant under the permutation of spins within each subgraph, which allows us to reduce the dimension of the Hilbert space under consideration [$\sim\mathcal{O}(N^2)$]. More details are provided in Appendix~\ref{sec:perm_invariance}. 

To test the NS-DQA paradigm, we adhere to Ref.~\cite{feinstein2024robustnessdiabatic} and consider the following nonstoquastic catalyst Hamiltonian,
\begin{equation}
    H_c = j_{xx} \sigma^x_i \sigma^x_j
\end{equation}
which constitutes a single coupling between any two spins (thanks to permutational invariance, the indices of the two spins are irrelevant) in $G_1$. 
As for the factor $j_{xx}$, this must be carefully chosen for each size to create a further maximally gap closing before the one characteristic of the MWIS annealing problem. To achieve this, a minimization process needs to be carried out, as already done in Refs.~\cite{feinstein2022xx,feinstein2024robustnessdiabatic}. 
 On the other hand, to test the SQS protocol, we first look for the optimal quench parameters $(B_{\text{q}}, \, \Delta T_{\text{q}}, \, t_{\text{q}})$ as described in Appendix~\ref{sec:sqs_optimization}.

\begin{figure}[t]
    \centering
    \includegraphics[width=\columnwidth]{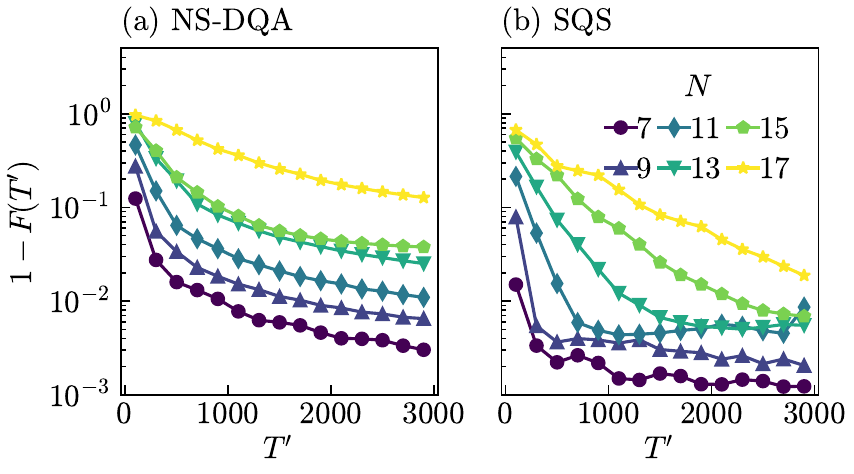}
    \caption{Infidelity $1 - F(T')$ as a function of the final annealing time $T'$ for different sizes and for the two different protocols. (a) NS-DQA. (b) SQS. Times are implicitly in units of $J^{-1}$.}
    \label{fig:unitary_scaling_fixed_size}
\end{figure}

Fig.~\ref{fig:comp_dqa_sqs_unitary} shows an example of dynamics of the ground-state fidelity in each of the two protocols (solid blue line) as a function of $t$ in comparison with the standard QA protocol (dotted green line). It is important to stress that in the SQS protocol, during the quench, the fidelity is always calculated with respect to the ground state of $H(t_\text{q})$ (i.\,e., the Hamiltonian right before the quench) and not with respect to the ground state of $H_{\text{q}}$ (i.\,e., the quenched Hamiltonian). 

In both protocols, the fidelity starts at $1$, since the system is prepared in the ground state at $t=0$. In the NS-DQA protocol, Fig.~\ref{fig:comp_dqa_sqs_unitary}(a), as soon as $t$ approaches the time of the first avoided crossing, the fidelity rapidly drops to zero, showing a sudden loss of overlap with the ground state. Only when the second avoided crossing is met does the fidelity macroscopically change again, and a high overlap with the ground state is restored. On the other hand, in the SQS protocol, shown in Fig.~\ref{fig:comp_dqa_sqs_unitary}(b), the initial value of the fidelity drastically changes when the quench starts, displaying periodic revivals. The oscillations reflect a more complex dynamical response of the SQS protocol compared to the monotonic decay in the NS-DQA case. %

In Fig.~\ref{fig:unitary_scaling_fixed_size}, we plot the final infidelity $1-F(T')$ of the two protocols as a function of $T'$ for different system sizes $N$. For the NS-DQA protocol, shown in panel~\ref{fig:unitary_scaling_fixed_size}(a), the infidelity decreases steadily with increasing $T'$ across all system sizes, without showing a clear saturation. This indicates that the NS-DQA protocol consistently benefits from longer annealing times. This is particularly interesting because, at long timescales, the adiabatic theorem guarantees convergence, while at short timescales, the diabatic protocol enhances it. These effects positively compete, ensuring that NS-DQA remains consistently effective and monotonically convergent. Conversely, the SQS protocol, panel~\ref{fig:unitary_scaling_fixed_size}(b), exhibits a much steeper decrease in infidelity for shorter annealing times, followed by saturation as $T'$ increases. This behavior suggests that the SQS protocol is more effective in rapidly reducing the infidelity, particularly for small $T'$, but then its performances saturate at long times and NS-DQA becomes the method of choice. Notably, for the sizes analyzed, the SQS protocol achieves significantly lower infidelity than the NS-DQA, with differences of up to one order of magnitude.

Furthermore, while standard quantum annealing leads to a degradation of fidelity that becomes increasingly pronounced with larger system sizes, due to the gap decreasing with $N$, as shown in Fig.~\ref{fig:fidelities_mwis}(a), diabatic annealing processes effectively address this issue, since the minimum gap plays a less crucial role in determining the algorithmic performance. Comparing the two behaviors reveals that, for the same annealing time, the fidelity in the standard annealing case is smaller by several orders of magnitude.

\begin{figure}[t]
    \centering
    \includegraphics[width=\columnwidth]{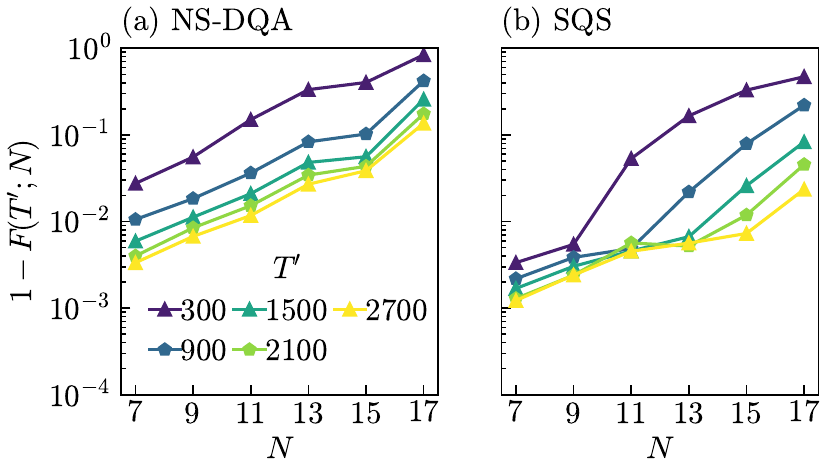}
    \caption{Infidelity $1 - F(T',N)$ versus the size of the system $N$, fixing the same duration of the dynamics for the two different protocols. (a) NS-DQA. (b) SQS. Times are implicitly in units of $J^{-1}$.}
    \label{fig:unitary_scaling_fix_tau}
\end{figure}

Finally, Fig.~\ref{fig:unitary_scaling_fix_tau} shows the scaling of the infidelity with the system size $N$, for fixed annealing times $T'$. In the NS-DQA protocol, panel~\ref{fig:unitary_scaling_fix_tau}(a), the infidelity exhibits a regular growth as $N$ increases, independently of the chosen $T'$. On the other hand, the SQS protocol, panel~\ref{fig:unitary_scaling_fix_tau}(b), shows a discontinuous growth in infidelity with $N$, with an effect of saturation that is more pronounced with higher annealing times.

\section{Dissipative dynamics}
\label{sec:dissipative}

The inclusion of dissipation is crucial for assessing the feasibility of annealing protocols on existing quantum devices, where interaction with the environment is inevitable. Understanding how dissipation influences the dynamics allows us to bridge the gap between idealized theoretical models and real-world quantum systems, shedding light on the robustness and performance of the protocol under nonideal conditions. Thus, in this section we consider a dissipative dynamics described by the master equation for the density matrix $\rho(t)$ of the system in the Lindblad form~\cite{lindblad1976equation, breuer2002openquantum}
\begin{align}
    \frac{\mathrm{d}}{\mathrm{d} t} \rho(t) &= -i\big[H,\rho(t)\big]+ 
    \nonumber \\
    &\phantom{=} 
    + \sum_\mu \Big( L_{\mu}\rho(t) L^{\dagger}_{\mu}
    -\frac{1}{2}\big\{L^{\dagger}_{\mu}L_{\mu},\rho(t)\big\} \Big) \;,
    \label{eq:lindblad}
\end{align}
where $H$ is the Hamiltonian of our two annealing protocols and the Lindblad operators $L_{\mu}$ characterize the dissipative dynamics. The Lindblad formalism provides a general framework to model the interaction of a quantum system with its environment under the assumptions of a Markovian bath and weak coupling. These approximations allow us to capture the essential features of dissipation while maintaining a tractable mathematical structure. This approach provides a framework to explore the impact of dissipation on the quantum dynamics of the system, offering insights into how it can influence fidelity and the effectiveness of different quantum protocols.

We focus on a specific type of dissipative model where the environment is represented by baths that preserve the system's permutation invariance, allowing us to exploit this symmetry for computational efficiency. This choice enables us to address realistic physical scenarios while maintaining a manageable level of complexity in our analysis.

Given the critical role of parameter selection in diabatic protocols, it is essential to verify whether the parameters optimized for unitary dynamics remain effective in the presence of dissipation. The influence of the environment can significantly modify the system's behavior, meaning that parameters ideal for unitary evolution may not necessarily yield the best results under dissipative dynamics. To simplify the analysis and focus on the core effects of dissipation, we use the same parameters optimized for the unitary protocols. This choice is supported by the analysis detailed in Appendix~\ref{sec:sqs_optimization}, which indicate that these parameters provide a good starting point for studying dissipative dynamics.

After this preliminary validation, we proceed to study the behavior of the infidelity $1-F(T')$ as a function of $T'$ under the effect of dissipation and repeat the comparison analysis performed in the unitary case. In particular, we explore the same range of annealing times and system sizes.

The timescale of dissipative effects is set by the relaxation time, which corresponds to the inverse of the eigenvalue with the smallest real part in the Lindbladian spectrum. In our case, it is of the order of $1/\gamma N$. Since we aim to investigate scaling effects with the system size, it is important to choose a dissipation rate $\gamma$ that depends on the size $N$ and ensures that dissipative effects are observable within the annealing times considered for all $N$. To achieve this, we introduce a reference time $T_{\scriptstyle{\mathrm{ref}}}$, setting the dissipation rate to $\gamma = 1 / (T_{\scriptstyle{\mathrm{ref}}} N)$, with $T_{\scriptstyle{\mathrm{ref}}}=1400$. This choice guarantees that the effects of dissipation become apparent for $t \sim T_{\scriptstyle{\mathrm{ref}}}$, independently of the system size, and that these effects are visible within the time window analyzed.

While more sophisticated master equations can more accurately describe the dissipative processes occurring in real devices—taking into account the structure of the energy levels of the system's Hamiltonian—we expect that, due to the presence of exponentially small energy gaps and the typical operating temperature of quantum annealers (typically around $\SI{10}{\milli\kelvin}$~\cite{king2017dwavetechnicalsheet}, equivalent to $\sim \SI{1.3}{\giga\hertz}$ in our units), all such descriptions will be qualitatively equivalent to simpler models assuming local dissipation, with no detailed information about the Hamiltonian of the reduced system.

In this context, we analyze the effects of two types of environments: dephasing and a gain-and-loss bath with emission and absorption channels whose rates satisfy detailed balance. The dephasing bath models processes that drive the system towards a maximally mixed state without relying on the system's specific energy spectrum. Similarly, the gain-and-loss bath effectively captures the interplay of emission and absorption processes that are expected to dominate in the presence of minimal energy gaps and finite temperatures. %

\subsection{Dephasing}
\label{subsec:dephasing}

\begin{figure}
    \centering
    \includegraphics[width=\columnwidth]{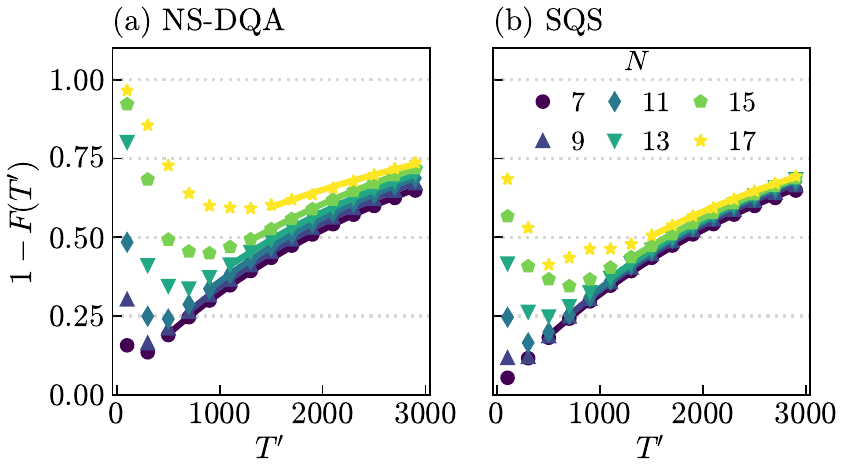}
    \caption{Infidelity $1 - F(T')$ vs annealing time $T'$ for the two different protocols with dephasing, plotted for increasing size $N$ of the system as markers and fit as solid lines of the same color. (a) NS-DQA protocol. (b) SQS protocol. Times are implicitly in units of $J^{-1}$.}
    \label{fig:deph_N}
\end{figure}

First, we consider a dephasing model, choosing Lindblad operators of the form $L_\mu=\sqrt{\gamma}\sigma^z_\mu$, where $\sigma^z$ is the Pauli-$Z$ operator, $\mu\in[1,N]$ is the site index, and $\gamma$ is the decay rate of the dissipative process. One can see that for this Lindbladian the steady state is the infinite-temperature thermal state $\rho = \mathds{1} / D$, where $D$ is the Hilbert space dimension and $\mathds{1}$ is the identity operator.  
In Fig.~\ref{fig:deph_N} we show how the infidelity depends on the final annealing time as the system size changes, for the NS-DQA protocol in panel (a) and SQS protocol in panel (b). In general, infidelities with dissipation are larger than the corresponding values obtained with unitary dynamics, given the same size and time $T'$, meaning that the dephasing mechanism is reducing the effectiveness of the two protocols studied. We can also see that, with the same dissipation, the SQS protocol still outperforms the NS-DQA protocol. Specifically, for short annealing times $T'$, the infidelity in NS-DQA increases with $N$ more rapidly than in SQS, which is less sensitive to the system size. For longer $T'$, the two protocols perform similarly, indicating that dephasing completely determines the evolution of the system. 

More specifically, for short dynamics (the shortest time considered is $T' = 100$), dissipation plays almost no role in the dynamics, and infidelity is determined by the same process that determined it in the unitary case, i.\,e., Landau-Zener transitions.
In fact, in that regime an exponential decrease with $T'$ of infidelity is observed, with a decay rate that depends on the minimum gap and thus changes with size: large sizes have a slower decay. 

At intermediate times, there is an interplay between coherent and dissipative regimes, resulting in the existence of an ‘optimal’ working annealing time $T'_\text{opt}$ that acts as a trade-off between LZ and dissipative processes~\cite{arceci2018optimal}. The optimal time $T'_\text{opt}$ increases with $N$ for both protocols, though more irregularly for the SQS protocol. This is likely a numerical effect due to the necessity of using different optimal parameters for every annealing time, as opposed to the optimal parameters used in NS-DQA, which only depend on $N$ and are fixed for all annealing times.

At longer times, dissipation dominates, and the expected trend is thermal relaxation, which causes the infidelity to increase to $(1-1/D) (1-e^{-T/T_1})$, where $D$ is the size of the Hilbert space and $T_1$ is the relaxation time. 
The asymptotic behaviors of the infidelity are fitted with the function 
\begin{equation}
    \label{eq:fit}
    \mathcal{I}(T') = y_0 + (y_\text{sat} - y_0)(1 - e^{-T'/\tau})
\end{equation}
where $y_0 = y(T'= 0)$ and $y_\text{sat} = y(T' \to \infty)$, which can be fixed to be $y_\text{sat} = 1$ disregarding the correction $-1/D$. The fits, performed for each size starting after the optimal working point $T'_\text{opt}$, are plotted as solid lines in Fig.~\ref{fig:deph_N} and we correctly recover that the saturation time $\tau$ is essentially size-independent (and protocol-independent). 

\begin{figure}
    \centering
    \includegraphics[width=\columnwidth]{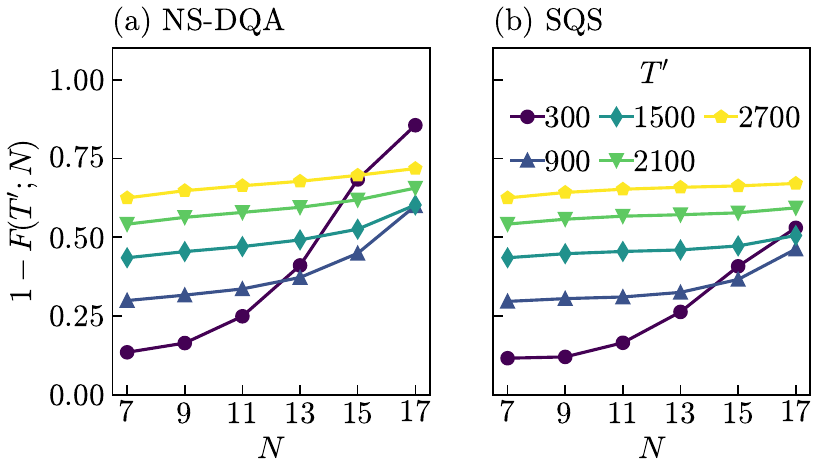}
    \caption{Infidelity $1 - F(T';N)$ vs size of the system $N$ for increasing annealing time $T'$ with dephasing. (a) NS-DQA protocol. (b) SQS protocol. Times are implicitly in units of $J^{-1}$.}
    \label{fig:deph_tau}
\end{figure}

In Fig.~\ref{fig:deph_tau} we show the scaling of the infidelity with the system size, for fixed annealing times $T'$, with dephasing, to compare the results with the one obtained with the unitary evolution shown in Fig.~\ref{fig:unitary_scaling_fix_tau}. For shorter times, it is evident that infidelity depends on the system size, because LZ transitions still play a role in the dynamics. At longer times, for dynamics $T' > 1500$, the dependence of the infidelity is dictated by the prefactor $(1 - 1/D)$ with $D$ being the Hilbert space dimension. %
In any case, looking at the differences between the two different regimes, unitary and dissipative, this type of dephasing process is capable of mitigating (even nullifying) the advantage gained through the implementation of diabatic protocols.

\subsection{Gain-and-loss bath}
\label{subsec:thermal_dissipation}
Finally, we consider two Lindblad channels in the form of absorption and emission, defined as $L^\uparrow_\mu=\sqrt{\gamma_\uparrow}\sigma^+_\mu=\sqrt{\gamma N_T}\sigma^+_\mu$ and $L^\downarrow_\mu=\sqrt{\gamma_\downarrow}\sigma^-_\mu=\sqrt{\gamma(N_T+1)}\sigma^-_\mu$, respectively. Here, $\sigma^\pm$ are the raising and lowering operators, $\mu \in [1,N]$ is the site index, and $\gamma$ is the same coefficient used in the dephasing protocol. The parameter $N_T={(e^{\beta \Omega_B}-1)}^{-1}$ is the thermal occupation number of the bosonic bath mode with characteristic frequency $\Omega_B = 1$, while $\beta$ is the effective inverse temperature of the bath. The gain and loss rates obey the detailed balance condition $\gamma_\uparrow/\gamma_\downarrow = N_T/(N_T+1) = e^{-\beta}$, ensuring that, at equilibrium, the excitation and de-excitation processes balance, with their ratio determined by the bath temperature.

It is important to note that this type of dissipation does not lead to the thermodynamic equilibrium of the reduced system's Hamiltonian. The detailed balance condition contains no information about the system's energy levels, and the dynamics are governed solely by the interaction with a thermal bath characterized by the effective temperature $\beta$. As such, this model only serves as an alternative form of dissipation to test the robustness and generality of the results obtained with dephasing by exploring a different type of environment.

\begin{figure}
    \centering
    \includegraphics[width=\columnwidth]{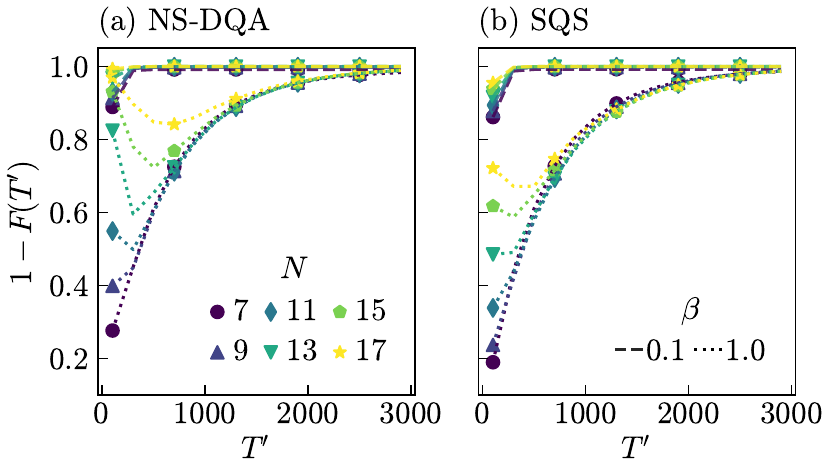}
    \caption{Infidelity $1 - F(T')$ as a function of the final annealing time $T'$ for different sizes and for the two different protocols with gain-and-loss bath with $\beta = 0.1, 1.0$. (a) NS-DQA. (b) SQS. Times are implicitly in units of $J^{-1}$.}
    \label{fig:thermal_scaling_fixed_size}
\end{figure}

At high bath temperatures, i.\,e., when $N_T \gg 1$, the system can reach highly excited states due to the dominance of gain processes. To analyze the effects of this type of dissipation, we fix $\beta=0.1$ and $\beta=1.0$, comparing the results, as done in the dephasing case, in Fig.~\ref{fig:thermal_scaling_fixed_size}. This allows us to assess whether the conclusions drawn in the presence of dephasing extend to environments with different dissipative mechanisms.

We present the trends of the infidelity using different line styles but maintaining the same color scale for each value of $\beta$. The trends are analyzed as a function of the annealing time in Fig.~\ref{fig:thermal_scaling_fixed_size}. The plots demonstrate that the smaller $\beta$, the more pronounced the destructive effect of gain-and-loss bath becomes, regardless of the final annealing time. For $\beta=1.0$, the behavior of the curves as $T'$ varies resembles what was previously observed in the case of dephasing: after an initial phase where the infidelity starts below one and continues to decrease, it reaches a minimum that depends on the chosen system size. Beyond this point, the infidelity begins to increase again, indicating the progressive failure of the annealing processes due to the influence of the environment. On the contrary, for $\beta=0.1$ (higher temperature of the bath) the curves are almost constant at $1$, signaling the complete failure of the annealing protocols. This can be understood considering that in this condition the system can reach excited states that prevent the system from diabatically reaching the instantaneous ground state at the end of the protocol.

\section{Conclusions}
\label{sec:conclusions}
In this paper, we explored and compared the effectiveness of two DQA protocols, NS-DQA and SQS, in solving a combinatorial optimization problem. These protocols leverage controlled excitations in the quantum system before it encounters a small energy gap. Specifically, we focused on the maximum weighted independent set problem.

First, we compared the NS-DQA protocol with the SQS approach. By optimizing parameters on a case-by-case basis, we verified that SQS consistently outperforms NS-DQA in terms of fidelity for shorter annealing times. However, NS-DQA demonstrated superior performance for longer durations due to its ability to effectively integrate adiabatic and diabatic dynamics.

Subsequently, we introduced two types of nonunitary dynamics: dephasing and gain-and-loss baths, to evaluate the robustness of the protocols in the presence of an external environment. For any annealing duration $T'$, the presence of noise significantly reduced the effectiveness of both protocols. In the case of dephasing, up to a critical time (which increases with system size), the performance of the protocols remained comparable to the unitary case. This suggests that for a fixed system size, a suitable compromise can be achieved between the bath strength and annealing duration without requiring experimentally unfeasible times. Beyond the critical time, however, the infidelity grows exponentially with $T'$, rendering optimization and annealing ineffective.

These results open avenues for applying diabatic protocols to real quantum devices. Future work should explore other STA approaches and evaluate the effectiveness of DQA protocols under different decoherence models. Additionally, extending the analysis to other optimization problems will help generalize the findings, allowing the integration of diabatic protocols into quantum optimization algorithms.

\begin{acknowledgments}
    G.\,P. and P.\,L. acknowledge financial support from PNRR MUR Project PE0000023-NQSTI and computational resources from MUR, PON “Ricerca e Innovazione 2014-2020”, under Grant No. PIR01\_00011 - (I.Bi.S.Co.). G.\,P.\ acknowledges computational resources from the CINECA award under the ISCRA initiative. This work was furthermore supported by the MUR project CN\_00000013-ICSC (P.\,L.), and by the QuantERA II Programme STAQS project that has received funding from the European Union’s Horizon 2020 research and innovation program.
\end{acknowledgments}

\appendix

\section{Hamiltonian parameters}\label{sec:app_hamiltonian}

In this Appendix, we derive the form of the maximum weighted independent set Hamiltonian, Eq.~\eqref{eq:ham_mwis}, and discuss the parameter setting of this model.

The MWIS problem aims to find the maximally weighted subset of vertices of a graph $G = (V, E)$, where each vertex in $V$ has a weight $\omega_i$, such that no vertices are connected by an edge in $E$. To express this problem in quadratic unconstrained binary optimization (QUBO) form, for each vertex we consider a binary variable $x_i \in \lbrace 0, 1 \rbrace $ such that, if $x_i = 1$, the vertex is included in the maximum independent set, and excluded otherwise. Then the goal is to minimize
\begin{equation}
    h_1(\vec{x}) = -\sum_{i\in V} \omega_i x_i
\end{equation}
with the constraint that no adjacent vertices are included in the independent set. Such constraint is expressed by the antiferromagnetic Hamiltonian
\begin{equation}
    h_2(\vec{x}) = \sum_{(i,j)\in E} x_i x_j,
\end{equation}
which adds a penalty every time two connected vertices are both included, $x_i = x_j = 1$.
The QUBO Hamiltonian representing the MWIS is thus given by
\begin{equation}
    H_\text{MWIS}(\vec{x}) = h_1(\vec{x}) + \lambda \, h_2(\vec{x}),
\end{equation}
where the parameter $\lambda > 0$ determines the strength of the constraint with respect to the first term. This parameter must be carefully tuned so that the penalty of including two adjacent spins in the set is higher than the energy gain given by $h_1$. To this end, a typical choice is to set $\lambda > 2 \max_i \omega_i$.

By applying the linear transformation $s_i = 2 x_i - 1 = \pm 1$ (such that vertex inclusion now corresponds to $s_i = +1$), it is possible to rewrite the MWIS Hamiltonian as
\begin{equation}
    H_\text{MWIS}(\vec{s}) = \text{const.} + \frac{1}{4}\sum_{i\in V} s_i (\lambda c_i - 2\omega_i) + \frac{\lambda}{4}\sum_{(i,j)\in E} s_i s_j,
\end{equation}
where $c_i$ denotes the connectivity of vertex $i$. Renaming $\lambda \to j_{zz}$, we see that this Hamiltonian corresponds, up to a constant, to Eq.~\eqref{eq:ham_mwis} upon rescaling the energy by $1/4$ and replacing $s_i \to \sigma_i^z$.

To select the optimal value of the parameter $j_{zz}$ such that the gap of the MWIS Hamiltonian closes near the end of the annealing process, it is necessary to adopt system-size-dependent precautions that account for the characteristic energy scales of both the problem Hamiltonian and the driving Hamiltonian. First, an initial value of the parameter, which we denote as $j_{zz}'=5.33$, is fixed. This value can remain unchanged as the system size varies. The latter primed anti-ferromagnetic coupling has to be normalized according to the characteristics of the system so that the energy scale of the problem and driving Hamiltonians result in having comparable energy scales \cite{feinstein2022xx}. 

At the same time, we preliminarily set the weights uniformly across the two subgraphs. We refer to $\omega_0'=W_0'/n_0$ for the sites in $G_0$ and $\omega_1'=W_1'/n_1$ for the sites in $G_1$, and $W_1'=1.0$, $W_0'=1.01$.

Once set these parameters, we normalize them so to eventually have 
\begin{equation}
    j_{zz}= e_{\textrm{scale}} K j_{zz}',
\end{equation}
and
\begin{equation}
    \omega_i= e_{\textrm{scale}} K \omega_i',
\end{equation}
where
\begin{equation}
    K=\frac{n_0 + n_1}{4 (n_0  n_1  j_{zz}' - 1)}, 
\end{equation}
and $e_{\textrm{scale}}=15$, which is size-independent. 

\section{SQS parameters optimization}\label{sec:sqs_optimization}

In case of the SQS protocol, for several given annealing times $T$ and system sizes $N$ we scanned through the parameter triple $(B_{\text{q}}, \, \Delta T_{\text{q}}, \, \tau_{\text{q}})$, where $\tau_{\text{q}} = t_{\text{q}}/T$, optimizing the final-state fidelity $F(\tau=1)$. In particular, varying $B_{\text{q}}$ and $ \tau_{\text{q}}$ in a grid of values, we look for the optimal value of $\Delta T_{\text{q}}$ that maximizes the final-state fidelity and store it. In this way we end up with two grid of values.

\begin{figure}[t]
    \centering
    \includegraphics[width=\columnwidth]{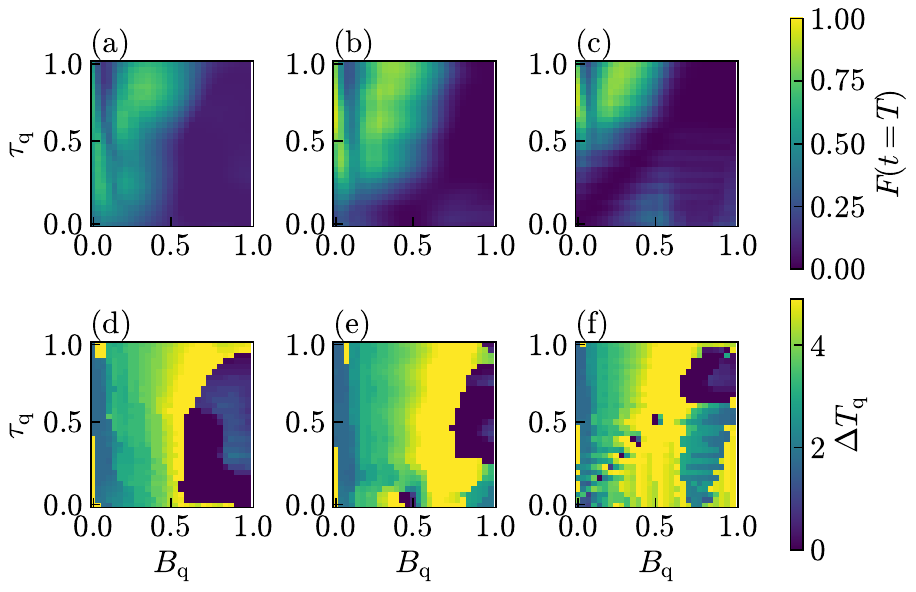}
    \caption{Grid search of the optimal final-state fidelity $F(t=T)$ for $N = 5$ for different sets of parameters $(B_{\text{q}}, \, t_{\text{q}}, \, \Delta T_{\text{q}})$. In the first row, we plot the fidelity for the best quench duration $\Delta T_{\text{q}}$ for each fixed combination of the first two parameters. This is done for three annealing times (a): $T=15$, (b):  $T=50$, (c): $T=100$. The corresponding optimal quench durations for each fixed set $(B_{\text{q}}, \tau_{\text{q}} )$ are shown in panels (d--f).}
    \label{fig:fid_heatmap}
\end{figure}

One example of this grid-search process can be visualized with the heat maps in Fig.~\ref{fig:fid_heatmap} for a system of $N=5$ qubits, where each column of plots corresponds to a different annealing time. In Figs.~\ref{fig:fid_heatmap}(a--c), we plot the fidelity for the best quench duration $\Delta T_{\text{q}}$ for each fixed combination of the first two parameters. This is done for three annealing times: $T=15$, $50$, $100$, respectively. The corresponding optimal quench durations for each fixed set $(B_{\text{q}}, \tau_{\text{q}} )$ are shown in Figs.~\ref{fig:fid_heatmap}(d--f). From the plot, it becomes clear that the maximum fidelity already saturates in the regime $T < 100$. It is interesting to note that the results of the optimization process give a value of $B_{\text{q}}$ that makes this SQS protocol substantially equivalent to a reverse-annealing protocol~\cite{perdomo2010studyheuristicguessesadiabatic, chancellor2017modernizingquantumannealing, passarelli2020reverse}.

\begin{figure}[t]
    \centering
    \includegraphics[width=\columnwidth]{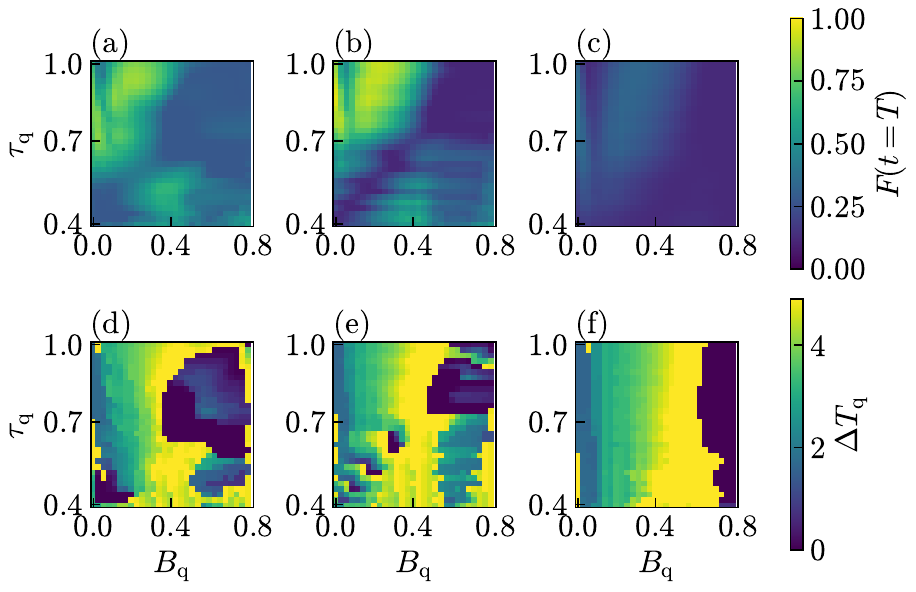}
    \caption{Grid search of the optimal final-state fidelity $F(t=T)$ for $N = 5$ for different sets of parameters $(B_{\text{q}}, \, t_{\text{q}}, \, \Delta T_{\text{q}})$ with the addition of dephasing with decay rate $\gamma=1/(NT_{\scriptstyle{\mathrm{ref}}})$, with $T_{\scriptstyle{\mathrm{ref}}}=50$.}
    \label{fig:fid_heatmap_dissipation_sz}
\end{figure}

Since dissipation could affect the optimization outcome obtained in the previous section, we repeat the optimization process in the presence of dephasing. The result is shown in Fig.~\ref{fig:fid_heatmap_dissipation_sz}, where we have used a decay rate of $\gamma = 1/(T_{\scriptstyle{\text{ref}}}N)$, where $T_{\text{ref}}=50$. 

Fig.~\ref{fig:fid_heatmap_dissipation_sz} shows that the qualitative aspect of the heat maps stays unaltered, thus leading to the possibility for us to continue considering the optimal parameters of the unitary dynamics, which are easier to retrieve. We have numerically verified that these results also holds for the other sizes analyzed in the manuscript.

\section{Permutational invariance}\label{sec:perm_invariance}

The permutational invariance of identical two-level particles (spins) enables an exponential reduction in the resources required to study the Lindbladian dynamics of spin and coupled boson ensembles, even when evolving under the influence of collective or local noise~\cite{nori2018permutation}. 

We can exploit the permutational symmetry of identical particles to simplify simulations by significantly reducing the Hilbert space dimension. For a system of $N$ spins, the Hilbert space is $2^N$-dimensional, scaling exponentially with the system size. This exponential growth makes it challenging to simulate large systems. However, by recognizing that, in each subgraph, $G_0$ and $G_1$, swapping two identical particles does not affect the physics, we can simplify the problem.

In our setup, the relevant permutation-invariant subgraphs are $G_0$, $G_1'$, and $G_c$. Here, $G_c$ contains the two sites connected by the catalyst, while $G_1'$ includes all sites of $G_1$ except the two linked by the catalyst.

As a result, the overall initial state can be expressed as a tensor product of three states, one for each subgraph, each symmetric under particle permutations within its respective subgraph
\begin{equation}
    \ket{\psi}=\ket{\psi_0}\otimes\ket{\psi_{1'}}\otimes\ket{\psi_c}.
\end{equation}

Given an $N$-spin state, we can always write it in the basis of the common eigenstates of the total spin $S^2$ and $S^z$ with quantum numbers $j\in[j_{\scriptstyle{\mathrm{min}}},N/2]$, $j_{\scriptstyle{\mathrm{min}}}=1/2$ and, $\forall j$, $m\in[-j,j]$ such that
\begin{equation}
\begin{split}
    &S^2 \ket{j,m} = j(j+1)\ket{j,m},\\
    &S^z\ket{j,m} = m\ket{j,m}.
\end{split}
\end{equation}
The latter is called Dicke states basis. It can be shown that the only symmetric states in this basis are those with $j=j_{\scriptstyle{\mathrm{max}}}=N/2$, namely the symmetric Dicke states~\cite{liu2015constructionsdickestateshigh}. If a system exhibits this symmetry, the analysis can be thus restricted to the subspace of the symmetric Dicke basis, i.\,e., that with fixed $j=j_{\scriptstyle{\mathrm{max}}}$~\cite{latorre2005entanglementpermutation}. In this case, the dimensionality of the Hilbert space is reduced from $2^N$ to $2j_{\scriptstyle{\mathrm{max}}}+1=N+1$ . Consequently, the problem no longer scales exponentially with $N$ but linearly, making it far more computationally efficient.

In our case, it is not possible to reduce the entire Hilbert space to a single $(N+1)$-dimensional space. Instead, it can be reduced to a space of dimension $(N_0+1)(N_{1'}+1)(N_c+1)=3(N_0+1)(N_1+1)=\frac{3}{4}(N^2-1)$, where $N_0$, $N_{1'}$, and $N_c$ correspond to the number of particles in the subgraphs $G_0$, $G_1'$, and $G_c$, respectively. Despite this, the reduction still provides a significant computational advantage, as the resulting dimensionality is far smaller than that of the full $2^N$-dimensional Hilbert space.  

\subsection{Unitary dynamics}
For each subgraph, all the collective spin operators of the model must then be constructed, for instance
\begin{equation}
    S_a^z=\frac{1}{2}\sum_{i\in G_a}\sigma_i^z,
\end{equation}
for whom the Dicke states $\ket{j_a,m_a}$ are eigenstates, and where $a=0,1',c$, and
\begin{equation}
    S_a^x=\frac{1}{2}\sum_{i\in G_a}\sigma_i^x.
\end{equation}
 In the first case, since Dicke states are eigenvectors of the collective operator $S_a^z$, the matrix elements of the latter are
 \begin{equation}
     \braket{j_a,m_a|S_a^z|j_a, m_{a'}}=m_a\delta_{m_a,m_{a'}},
 \end{equation}
where the quantum number $j_a$ is fixed due to the permutational symmetry, i.\,e., the only allowed transitions are in the $J^z$ subspace with maximum quantum number.
In the second case, we can write $S_a^x$ in terms of raising and lowering operators, so to have
\begin{equation}
\begin{split}
    &\braket{j_a,m_a|S_a^x|j_a,m_{a'}}=\frac{1}{2}\braket{j_a,m_a|\left(S_a^++S_a^-\right)|j_a,m_{a'}}=\\
    &=\frac{\delta_{m_{a'},m_a+1}}{2}\sqrt{(j_a-m_a)(j_a+m_a+1)}+\\
    &+\frac{\delta_{m_{a'},m_a-1}}{2}\sqrt{(j_a+m_a)(j_a-m_a+1)}.
\end{split}
\end{equation}
Hence, we can first rewrite the Hamiltonian of the MWIS problem of Eq.\eqref{eq:ham_mwis} in the form~\cite{feinstein2022xx, feinstein2024robustnessdiabatic}
\begin{equation}
\begin{split}
    H_p &= \left(n_1 j_{zz}-2\frac{W_0}{n_0}\right)\sum_{i \in G_0}\sigma_i^z+\\
    &+\left(n_0 j_{zz}-2\frac{W_1}{n_1}\right)\sum_{i\in G_1}\sigma_i^z+\\
    &+j_{zz}\sum_{(i,j)\in{\text{edges}}}\sigma_i^z\sigma_j^z,
\end{split}
\end{equation}
and then, in terms of collective operators, as
\begin{equation}
    H_z = 2 h_0 S_0^z + 2 h_1 \left(S_{1'}^{z}+S_c^z\right)+4j_{zz}S_0^z\left(S_{1'}^z+S_c^z\right), 
\end{equation}
where $h_0 = n_1 j_{zz}-2 W_0/n_0$, $h_1=n_0 j_{zz}-2 W_1/n_1$. At the same time, we can rewrite
\begin{equation}
    H_x=-\sum_{i\in G_0,G_{1'},G_c}\sigma_i^x=-2\left(S_0^x+S_{1'}^x+S_c^x\right)
\end{equation}
and
\begin{equation}
    H_c=j_{xx}\sigma_i^x \sigma_j^x, \hspace{2mm} (i,j)\in G_c
\end{equation}
in the collective counterpart considering that
\begin{equation}
    S^{x2}=\frac{1}{4}(\sigma_i^x\sigma_j^x+\sigma_j^x\sigma_i^x+\sigma_i^x\sigma_i^x+\sigma_j^x\sigma_j^x),
\end{equation}
and
\begin{equation}
    4 S^{x2} - 2\mathds{1}=\sigma_i^x\sigma_j^x+\sigma_j^x\sigma_i^x= 2\sigma_i^x\sigma_j^x,
\end{equation}
so that
\begin{equation}
    H_c =j_{xx}\left(2S_c^{x2}-\mathds{1}\right).
\end{equation}

\subsection{Dissipative dynamics}
The challenge in dealing with a Lindbladian with local operators lies in the fact that, in general, their action on Dicke states is not straightforward. As a result, the Lindblad equation with local operators does not restrict its dynamics to the subspace of $N+1$ symmetric states. However, it still preserves permutational invariance, which allows for a significant simplification~\cite{nori2018permutation}.

Consider a density matrix $\rho$ that is initially built to be symmetric under permutations. In general, $\rho$ can be projected onto the $\ket{j,m}\bra{j',m'}$ basis of Dicke states. It can then be shown that the Lindblad equation does not generate coherences between $m$ and $m'$ for different $j$ values, although it allows the system to explore subspaces with $j$ values other than the maximum.

Thus, $\rho$ can be expressed as:
\begin{equation}
    \rho = \bigoplus_j \rho_j = \sum_j p_{jmm'}\ket{j,m}\bra{j,m'},
\end{equation}
where $p_{jmm'}=\braket{j,m|\rho|j,m'}$. Thus, $\rho$ can be written as a block-diagonal matrix, where the off-diagonal blocks are inaccessible due to permutational symmetry. Each diagonal block corresponds to a fixed $j$ subspace and is further characterized by its degeneracy. Specifically, for a given $j$, the degeneracy is~\cite{nori2018permutation}
\begin{equation}
    d_N^j= (2j+1)\frac{N!}{(N/2+j+1)!(N/2-j)!},
\end{equation}
that grows as $j$ decreases. The total number of elements of the density matrix is
\begin{equation}
    \sum_{j=j_{\scriptstyle{\mathrm{min}}}}^{N/2}=\frac{1}{6}(N+1)(N+2)(N+3)\sim \mathcal{O}(N^3),
\end{equation}
so that there is still a computational advantage.

In particular, we are interested in local dephasing, which can assume the general form 
\begin{equation}
    \gamma\sum_{\mu=1}^N\mathcal{L}_{\sigma^z_\mu}[\rho] = 4\gamma \left[\left(\sum_{\mu=1}^NS^z_\mu \rho S^{z\dagger}_\mu\right)-\frac{N}{2}\rho\right],
\end{equation}
so that the Lindblad equation for the matrix element of the density matrix becomes
\begin{align}
    &\dot{p}_{jmm'}(t)\ket{j,m}\bra{j,m'}=p_{jmm'}\gamma\\
    &\quad\times\left[2\left(\sum_{\mu=1}^NS^z_\mu\ket{j,m}\bra{j,m'}S^{z\dagger}_\mu\right)-\frac{N}{2}\ket{j,m}\bra{j,m'}\right].\notag
\end{align}
In Refs.~\cite{chase2008collectiveprocesses, baragiola2010collectiveuncertainty} it is eventually shown that, under the effect of pure local dephasing, the master equation of the density matrix assumes the form
\begin{equation}
\begin{split}
    &\dot{p}_{jmm'}\ket{j,m}\bra{j,m'}=\gamma\bigg[\left(\frac{N}{2}-mm'\frac{N/2+1}{j(j+1)}\right)\\
    &\times\ket{j,m}\bra{j,m'}+B_z^{jm}B_z^{jm'}\frac{N/2+j+1}{j(2j+1)}\\
    &\times\ket{j-1,m}\bra{j-1,m'}+D_z^{jm}D_z^{jm'}\frac{N/2-j}{(j+1)(2j+1)}\\
    &\times\ket{j+1,m}\bra{j+1,m}\bigg],
\end{split}
\end{equation}
Where $B_z^{jm}=\sqrt{(j+m)(j-m)}$ and $D_z^{jm}=\sqrt{(j+m+1)(j-m+1)}$. 

The same reasonings can be repeated for the local gain-and-loss bath considering local channels of absorption and emission. For the explicit calculations we remind to Ref.~\cite{nori2018permutation}. All results shown in the main text were obtained using tools provided by the Piqs library, which is now part of the Python QuTiP package~\cite{lambert2024qutip5, nori2018permutation}.

\end{document}